# Second Harmonic Generation with 48% Conversion Efficiency from Cavity Polygon Modes in a Monocrystalline Lithium Niobate Microdisk Resonator

Chao Sun[1,2,#], Jielei Ni[3,#], Chuntao Li[1,2], Jintian Lin[4,5,7,†], Renhong Gao[1,‡], Jianglin Guan[1,2], Qian Qiao[4,6], Qifeng Hou[4,7], Xiaochao Luo[4,5], Xinzhi Zheng[1,2], Lingling Qiao[4], Min Wang[1], and Ya Cheng[1,2,4,6,8,9,10,*]

[1]*The Extreme Optoelectromechanics Laboratory (XXL), School of Physics and Electronic Science, East China Normal University, Shanghai 200241, China*
[2]*State Key Laboratory of Precision Spectroscopy, East China Normal University, Shanghai 200062, China*
[3]*Nanophotonics Research Center, Institute of Microscale Optoelectronics & State Key Laboratory of Radio Frequency Heterogeneous Integration, Shenzhen University, Shenzhen 518060, China*
[4]*State Key Laboratory of High Field Laser Physics and CAS Center for Excellence in Ultra-Intense Laser Science, Shanghai Institute of Optics and Fine Mechanics (SIOM), Chinese Academy of Sciences (CAS), Shanghai 201800, China*
[5]*Center of Materials Science and Optoelectronics Engineering, University of Chinese Academy of Sciences, Beijing 100049, China*
[6]*School of Physical Science and Technology, ShanghaiTech University, Shanghai 200031, China*
[7]*School of Physical Sciences, University of Science and Technology of China, Hefei 230026, China*
[8]*Shanghai Research Center for Quantum Sciences, Shanghai 201315, China Hefei National Laboratory, Hefei 230088, China*
[9]*Collaborative Innovation Center of Extreme Optics, Shanxi University, Taiyuan 030006, China*
[10]*Collaborative Innovation Center of Light Manipulations and Applications, Shandong Normal University, Jinan 250358, China*
[#]*These authors contributed equally to this work*

[†]*jintianlin@siom.ac.cn*
[‡]*rhgao@phy.ecnu.edu.cn*
*ya.cheng@siom.ac.cn

Thin-film lithium niobate (TFLN) based optical microresonators offer large nonlinear coefficient $d_{33}$ and high light-wave confinement, allowing highly efficient second-order optical nonlinear frequency conversion. Here, we achieved ultra-efficiency SHG from high-Q polygon modes by maximizing the utilization of the highest nonlinear coefficient $d_{33}$ in a monocrystalline X-cut TFLN microdisk resonator for the first time. The polygon modes are designed and formed with two parallel sides perpendicular to the optical axis of the lithium niobate crystal by introducing weak perturbations into the microdisk of a tapered fiber, which maximizes the utilization of $d_{33}$. The polygon modes exhibit ultrahigh intrinsic Q factors of ~$3.86\times10^7$, due to the fact that polygon modes are located far from the relatively rough sidewall of the microdisk. Moreover, the pump and second harmonic polygon modes share high modal overlap factor of ~80%. Consequently, SHG from cavity polygon modes with absolute conversion efficiency as high as 48.08% was realized at an on-chip pump level of only 4.599 mW without fine domain structures, surpassing the best results (23% and 30%) reported in other two domain-inversion-free phase matching schemes and even approaching the record (52%) in PPLN microresonators.

## 1. Introduction

Photonic integrated highly efficient second harmonic generation (SHG) frequency convertors enable systems-on-chip solutions for a wide range of applications including precision frequency metrology, optical clocks, molecular imaging, and quantum-information processing, by leveraging quadratic nonlinearity ($\chi^{(2)}$)[1-6]. Translating the SHG convertors from benchtop to integrated platforms will bring high conversion efficiency with low pump levels, lower cost, compact footprint, and robustness. Thin-film lithium niobate (TFLN) microresonators are particularly attractive for on-chip SHG, due to its significant $\chi^{(2)}$ ($d_{33}$~41.7 pm/V), wide transparency window from 0.35 to 5.0 μm, and strong optical field confinement[7-13]. To boost the conversion efficiency of SHG, several prerequisites must be satisfied simultaneously consisting of the utilization of the highest $\chi^{(2)}$ coefficient of $d_{33}$, the fulfillment of phase match and double-resonance conditions with ultra-high Q modes of large spatial modal overlap factor[14-16]. The flexibility in ferroelectric microdomain inversion enables quasi-phase-matched PPLN photonic microresonators from fundamental whispering gallery modes (WGMs), showing extremely high nonlinear optical conversion efficiencies by leveraging $d_{33}$[1,6,17]. However, the domain inversion severely increases the fabrication complexity and scattering loss in the domain wall[7,17]. Alternatively, two domain-inversion-free phase matching schemes have been developed partially anticipating with high-order WGMs, including novel mode-phase-matching (MPM) method using reverse-polarized double-layer X-cut TFLN microresonators[2,12,18], and natural quasi-phase matching scheme using X-cut monocrystalline microdisks[15,19]. Despite these advances, success in boosting the conversion efficiency is still extremely restricted by the accessible low Q factor due to the difficulty in quality fabrication of the dual-layer microresonators[2,12] or the limited modal overlap factor (<10%) because of the use of high-order WGMs and the spatial utilization of $d_{33}$[15,19], respectively.

Here, in contrast to using WGMs, we demonstrate a unique ultra-efficiency second harmonic generation (SHG) mechanism from cavity polygon modes in an X-cut dispersion engineered TFLN microdisk without fine microdomain structures, by maximum leveraging the largest nonlinear coefficient $d_{33}$ and the high modal overlap factor. Specific polygon modes with a spatial square-pattern profile of two parallel sides perpendicular to the optical axis of lithium niobate crystal is intentionally excited by introducing weak perturbation with a coupled tapered fiber, leading to the maximum utilization of $d_{33}$. Meanwhile, the polygon modes exhibit ultrahigh intrinsic Q factors of ~3.86×10$^7$, due to the fact that polygon modes are located far from the relatively rough sidewall of the microdisk[20-22]. Moreover, the pump and second harmonic polygon modes share high modal overlap factor of ~80%[20,21]. Consequently, by pumping the polygon mode at 1560 nm, natural quasi-phase matched SHG with an absolute conversion efficiency of 48.08% was realized at an on-chip pump level of only 4.599 mW without fine domain structures, approaching the best previously reported result (52%) in PPLN microresonators[23].

## 2. Natural quasi-phase matched SHG with ultra-efficiency from polygon modes

### 2.1 Fabrication of the TFLN microdisks

The TFLN microdisk resonator was fabricated on a commercially available (NanoLN) 700-nm-thick X-cut TFLN wafer by femtosecond laser lithography assisted chemo-mechanical etching (PLACE) technique with five steps[24]. First, a layer of chromium (Cr) was deposited on the TFLN wafer. The Cr layer was then subtractively ablated into microdisk patterns with a high spatial resolution of 200 nm by focused femtosecond laser. Subsequently, the microdisk patterns were transferred from the Cr layer to the lithium niobate thin film through chemo-mechanical polishing (CMP). Fourth, the Cr patterns were removed away by chemical etching and secondary CMP was carried put to polish the surface of the fabricated TFLN microdisks, and reduce the thickness of the microdisks to 0.59 μm. Finally, the silicon dioxide layer beneath the TFLN microdisks was undercut as small pedestals to support the suspended TFLN microdisks. The diameter of the microdisk was designed and fabricated as 57.28 μm, as shown in the inset of Fig. 1, and the wedge angle of the sidewall of the microdisk was 21°.

### 2.2 The experimental setup

The experimental setup is schematically depicted in Fig. 1. A narrow linewidth tunable diode laser (Model: TLB-6728, New Focus Inc.) was used as the pump light source. The pump light was selectively tuned to be transverse-electrically polarized using an online fiber polarization controller. A variable optical attenuator (VOA) was employed to control the pump power. A tapered fiber with a waist diameter of 2 µm was fabricated by thermally pulling a section of standard single-mode optical fiber to couple with the microdisk. The tapered fiber was positioned in direct contact with on the top surface of the microdisk at an angle of 45° between the tapered fiber and the optical axis of the lithium niobate. The distance from the microdisk center to the tapered fiber was set as 27.32 µm, in order to simultaneously introduce weak perturbation into the microdisk, couple the pump light into the microdisk, and couple the generated SHG signal out of the microdisk. An optical imaging system consisted of a microscope objective with a numerical aperture (NA) of 0.42, a visible charge coupled device (CCD) camera or an infrared (IR) charge-coupled device (CCD) was mounted above the microdisk to real-time monitor the microdisk-fiber-coupling system. The IR CCD was used to capture the intensity profile of the pump light in the microdisk, while the visible CCD was used to capture the intensity profile of the SHG scattering from the top surface of the microdisk. The output signals were sent into a photodetector (PD, Model: 1611-FC, New Focus, Inc.) to monitor the output power, a power meter to measure the output power, or an optical spectrum analyzer (OSA, YOKOGAWA, Inc., Model AQ6370D) to analyze the optical spectrum. A short-pass filter (model: FESH800, Thorlabs, Inc.) and a long-pass filter (model: FELH800, Thorlabs, Inc.) were used to block the pump light and the SHG signal, when the output power of the SHG and the input power of the pump light were measured, respectively. For measuring the Q factors of the modes, a signal generator was used to linearly scan the wavelength of the tunable diode laser.

An oscilloscope (Model: MDO3104, Tektronix Inc.), connected to the photodetector, was used to record the transmission spectrum during wavelength scanning.

**2.3 Ultra-efficient SHG**

When the pump laser wavelength was set to 1560.6 nm, a strong SHG signal was observed at a wavelength of 780.3 nm. Figures 2(a) and 2(b) show the spectra of the pump light and the SHG signal, respectively. The captured intensity profile of the SHG signal, which scatters from the top surface of the microdisk, shows a bright square pattern, as illustrated in the inset of Fig. 2(b). The corresponding intensity profile of the pump light scattering from the top surface of the microdisk also exhibits a square pattern, as shown in the inset of Fig. 2(a). It is worth noting the formation of such polygon modes relies on mode recombination of quasi-degeneracy of WGMs induced by weak perturbation[25,26]. And the weak perturbation mostly results in the exchange of components between adjacent eigenmodes with geometrical features closest to each other. Therefore, the modal overlap factor among these polygon modes within the same square mode family reaches up to 86%[19-21], despite a broad spectral wavelength range. This high modal overlap will significantly boost the conversion efficiency of the SHG. When the pump power grows, the output power of the SHG signal rapidly increases, as shown in Fig. 2(c). Furthermore, the SHG conversion efficiency linearly grows with the increasing pump power, as shown in Fig. 2(d), which agrees well with the nature of the SHG process. From the linear fitting, the normalized conversion efficiency is determined to be 7.5%/mW. It is important to note that, this value is underestimated since only the power of the SHG signal output from the tapered fiber was collected with a low coupling efficiency. When the on-chip pump power was further increased to 4.599 mW, the power of the SHG signal was measured to be 2.211 mW, as shown in Figs 2(a) and 2(b). Therefore, the absolute conversion efficiency of SHG is determined to 48.08%. This value surpasses the best results (23% and 30%) reported in the two domain-inversion-free phase matching schemes[2,13,19], and even approaches the best results (52%) recorded in quasi-phase-matched PPLN photonic microresonators[23].

**2.4 Characterization of the Q factors and calculation of the effective refractive indices of the polygon modes**

It is necessary to reveal the mechanism underlying the ultra-efficiency SHG. First, besides of the high modal overlap, the Q factors of the modes play important poles in the enhancement of the nonlinear optical interplay, for building up the light field. The loaded Q factor of the pump mode around 1560.6 nm was measured to be $2.69 \times 10^6$, as shown in Fig. 3(a). The transmittance was only 74.1%. As a result, the intrinsic Q factor was calculated to be $3.86 \times 10^7$. Meanwhile, the second harmonic mode around 780.3 nm also displayed a high loaded Q factor, which was measured to be $1.02 \times 10^6$, as shown in Fig. 3(b). And the intrinsic Q factor was determined to be $3.43 \times 10^7$. These Q factors of the second harmonic mode represent the best result in the 775 nm waveband reported in TFLN platform, which will remarkably promote the SHG efficiency.

Due to the fact that these square modes are primarily located within the microdisk, which is typically situated far from the cavity periphery, the effective refractive indices of these square modes are equivalent to those of the fundamental modes in the one-dimensional (1D) slab TFLN waveguide. Considering that the thickness of the TFLN microdisk is ~ 590 nm, the effective refractive indices for the pump polygon mode and the second harmonic polygon mode were calculated to be 1.9746 and 2.1504, respectively. Therefore, the phase mismatch $\Delta k=2k_F\text{-}k_{SHG}$ is $1.4159\times10^6$ (m$^{-1}$). Here, $k_F$ and $k_{SHG}$ are wave vectors of the pump light and the SHG signal, respectively.

**2.5 The natural quasi-phase match scheme based on polygon modes**

To boost the SHG efficiency, natural nonlinear "poling" based on the polygon modes in the X-cut TFLN microdisk provides additional optical momentum to compensate the phase mismatch. Because the modes participating the SHG process in the X-cut TFLN microdisk are transverse-electrically polarized, the second-order nonlinear coefficient can be expressed as[15]

$$d_{eff} = -d_{22}cos^3\alpha + 3d_{31}cos^2\alpha sin\alpha + d_{33}sin^3\alpha, \quad (1)$$

where $\alpha$ is the angle between the wave vector $\vec{k}$ of the TE-polarized wave and the optical axis. Assuming that the light field within the square pattern is uniform, the angle $\alpha$ changes four times as light travels one circle in the cavity. The dependence of the nonlinear coefficient $d_{eff}$ on the azimuth angle $\theta$ is plotted in Fig. 4(a). The nonlinear coefficient $d_{eff}$ equals four discrete values, which are -2.46 pm/V (i.e., -$d_{22}$) at $\theta = $ -

π/4~π/4, -41.7 pm/V (i.e., $d_{33}$) at $\theta$=π/4~3π/4, 2.46 pm/V (i.e., $d_{22}$) at $\theta$= 3π/4~5π/4, 41.7 pm/V (i.e., -$d_{33}$) at $\theta$=5π/4~7π/4, respectively. This dependence clearly shows that although $d_{eff}$ processes two amplitudes, the sign of $d_{eff}$ inverts every half cycle in the TFLN microdisk as illustrated in Fig. 4(a). This is analog to the periodic microdomain inversion in a PPLN crystal, leading to an additional momentum for compensating the phase mismatch to boost the SHG efficiency.

Under slowly varying amplitude approximation, the growth rate of the field amplitude $E_{SHG}$ of the SHG signal at the propagated path $z$ along the square pattern can be expressed as

$$E_{SHG}(z) = \frac{2id_{eff}\omega_{SHG}^2 E_P^2}{k_{SHG}c^2} \int_0^L e^{i\Delta k(z) \cdot z} dz. \qquad (2)$$

Here, $\omega_{SHG}$, $E_P$, and $c$ are the angular frequency of the SHG signal, the field amplitude of the pump light, and the light speed in vacuum. The calculated field amplitude $|E_{SHG}|$ as light propagates along two cavity cycles in the microdisk is shown in Fig. 4(b). It is necessary to carefully analyze how the field amplitude $|E_{SHG}|$ grows within each half cycle in the microdisk. At $\theta$ ranging from -π/4 to π/4 (or from 7π/4 to 9π/4), since the nonlinear coefficient is relatively small, $|E_{SHG}|$ oscillates with ~18.5 periods (i.e., 37 coherent lengths) and there is a very small gain. At $\theta$ ranging from π/4 to 3π/4 (or from 9π/4 to 11π/4), although $|E_{SHG}|$ also oscillates with ~18 periods, there is a remarkable gain because of the utilization of the largest nonlinear coefficient of $d_{33}$. Because the sign of $d_{eff}$ inverts every half cycle in the TFLN microdisk, at the next half cycle in the microdisk, there is also a remarkable gain as the front half cycle, akin to the growth rate in the each coherent length of PPLN counterpart. Therefore, thanks to the high Q factors, the high modal overlap factor, the specific natural quasi-phase match based on polygon modes, ultra-efficiency SHG signal is demonstrated.

## 3. Conclusion

In summary, we have formed high-Q polygon optical modes with a square pattern profile of two parallel sides perpendicular to the crystal optical axis in the TFLN microdisk for the first time to maximize the utilization of $d_{33}$ and the modal overlap, showcasing ultra-efficiency SHG with a high conversion efficiency upto 48%. These observations open new avenues for exploring on-chip efficient classical and quantum nonlinear light sources.

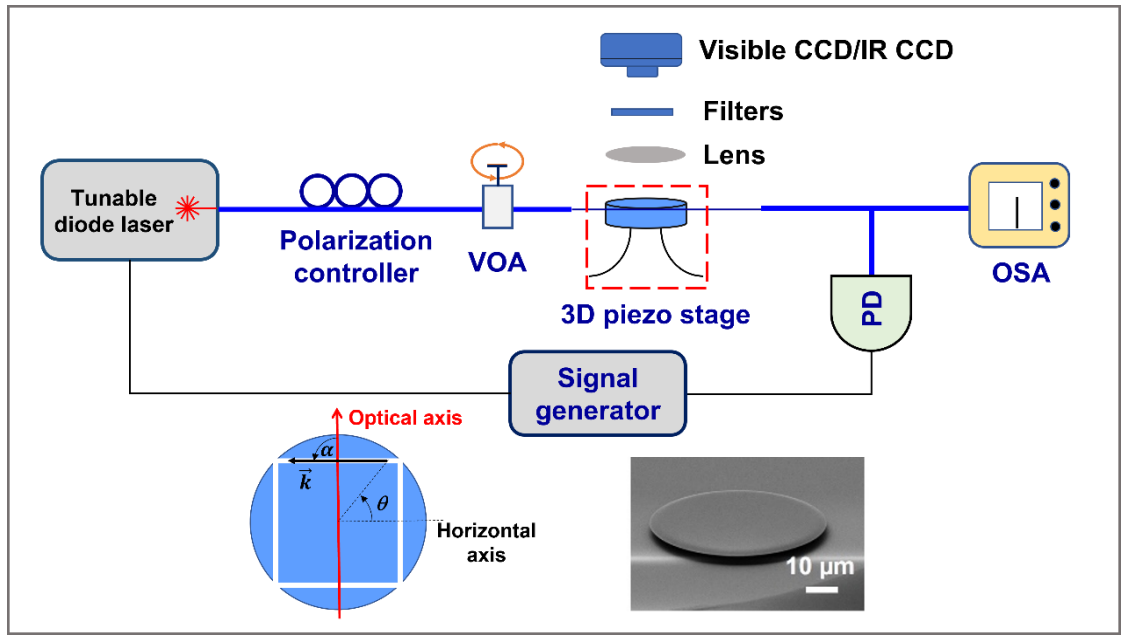

**Figure 1.** The experimental setup for SHG. The lower left inset and lower right inset show the exitation of square modes in the X-cut TFLN microdisk and the SEM image of the TFLN microdisk.

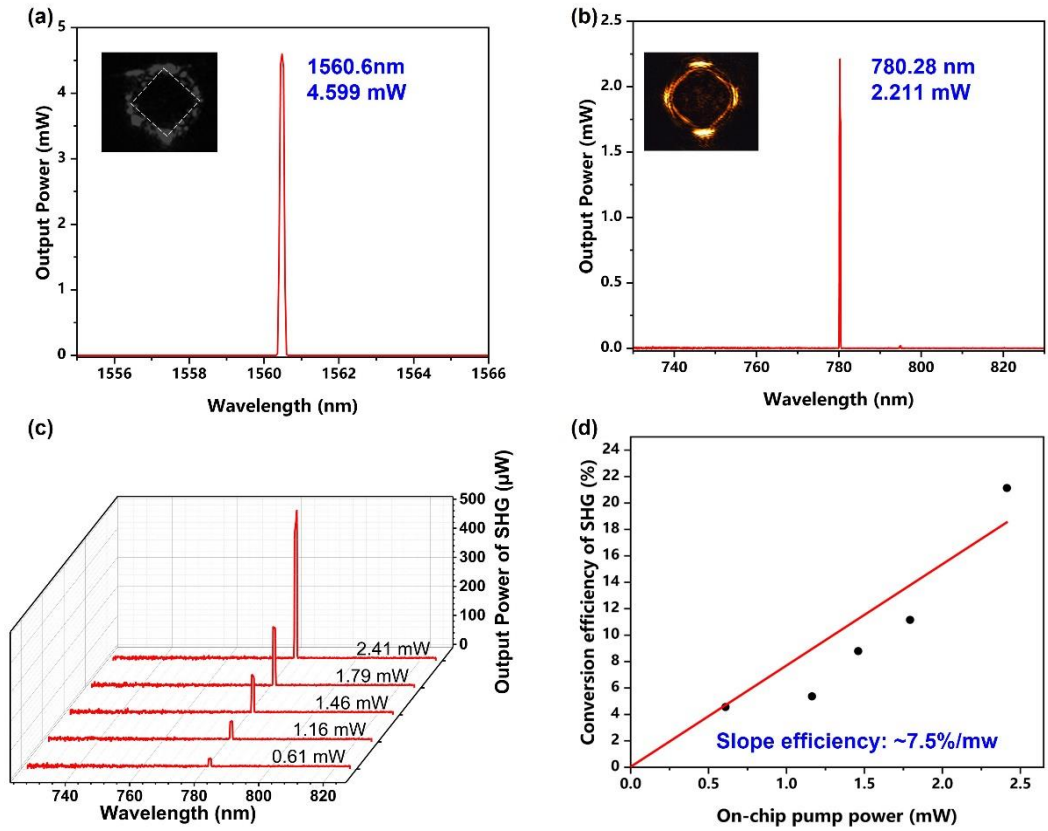

**Figure 2.** (a) Spectrum of the pump laser at 1560.60 nm wavelength. Inset: the optical micrograph of the pump light, showing a square-pattern intensity profile. (b) Spectrum of the second harmonic generated at 780.30 nm. Inset: the optical micrograph of SHG signal, also exhibiting a square-pattern intensity profile. (c) SHG output power varied with on-chip pump powers. (d) The SHG conversion efficiency as a function of the on-chip pump power.
15

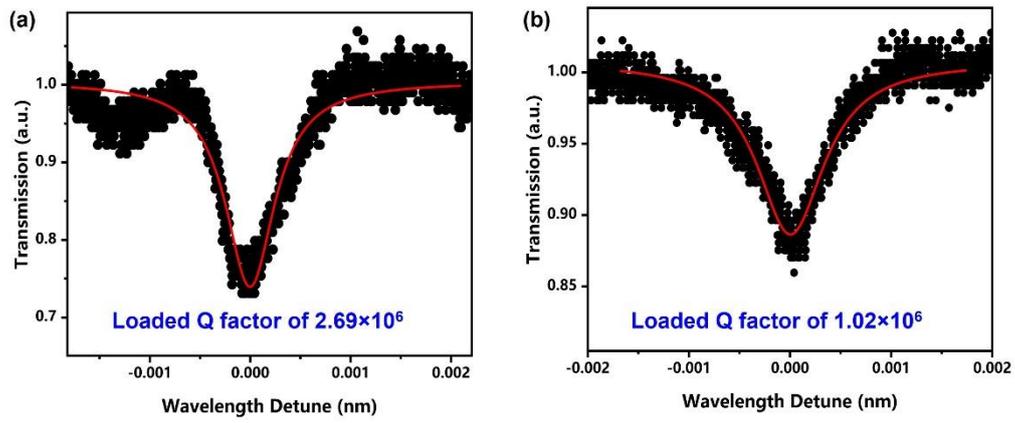

**Figure 3.** (a) The loaded Q factor of the pump mode. (b) The Q factor of the second harmonic mode.

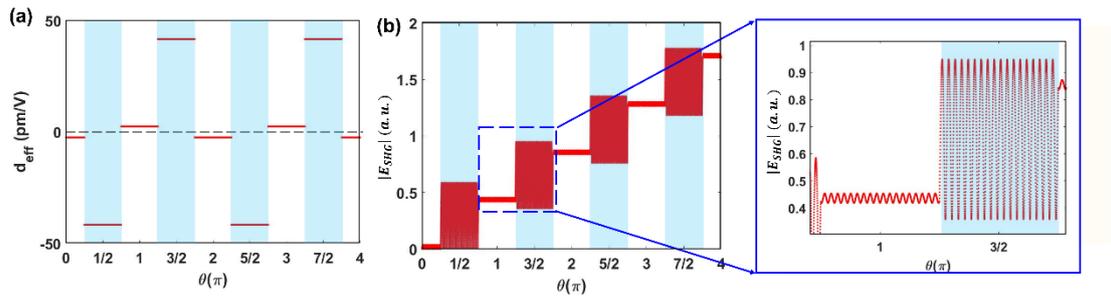

**Figure 4**. (a) The dependence of the nonlinear coefficient $d_{\text{eff}}$ on the azimuth angle $\theta$. (b) The growth of $|E_{\text{SHG}}|$ versus $\theta$ when the light wave propagates along 2 cavity cycle.